\documentclass[twocolumn,showpacs,preprintnumbers,amsmath,amssymb]{revtex4}

\usepackage{graphicx}
\usepackage{graphicx,epstopdf}
\usepackage{xcolor}
\usepackage{dcolumn}
\usepackage{bm}
\usepackage{mathrsfs} 
\usepackage{mathtools}
\usepackage{hyperref}

\begin{document}

\title{Amplitude mediated spiral chimera pattern in a nonlinear reaction-diffusion system}
\author{Srilena Kundu$^1$}
\author{Paulsamy Muruganandam$^2$}
\author{Dibakar Ghosh$^1$}
\email{diba.ghosh@gmail.com}
\author{M. Lakshmanan$^3$}
\affiliation{$^1$Physics and Applied Mathematics Unit, Indian Statistical Institute, 203 B. T. Road, Kolkata-700108, India\\
$^2$Department of Physics, Bharathidasan University, Tiruchirapalli-620024, India \\
$^3$Department of Nonlinear Dynamics, School of Physics, Bharathidasan University, Tiruchirapalli-620024, India}

\date{\today}

\begin{abstract}
Formation of diverse patterns in spatially extended reaction-diffusion systems is an important
aspect of study which is pertinent to many chemical and biological processes. Of special interest
is the peculiar phenomenon of chimera state having spatial coexistence of coherent and incoherent
dynamics in a system of identically interacting individuals. In the present article, we report the
emergence of various collective dynamical patterns while considering a system of prey-predator
dynamics in presence of a two-dimensional diffusive environment. Particularly, we explore the
observance of four distinct categories of spatial arrangements among the species, namely spiral
wave, spiral chimera, completely synchronized oscillations, and oscillation death states in a broad
region of the diffusion-driven parameter space. Emergence of amplitude mediated spiral chimera states displaying
drifted amplitudes and phases in the incoherent subpopulation is detected for parameter values beyond both Turing and Hopf bifurcations. Transition scenarios among all these
distinguishable patterns are numerically demonstrated for a wide range of the diffusion coefficients which reveal that the chimera states arise during the transition from oscillatory to steady state
dynamics. Furthermore, we characterize the occurrence of each of the recognizable patterns by
estimating the strength of incoherent subpopulations in the two-dimensional space.
\end{abstract}


\maketitle


\section{Introduction}\label{intro}

Pattern formation in reaction-diffusion (RD) systems has been a crucial area of research since ages for chemists, biologists and physicists due to its substantial applications in understanding the processes associated with morphogenesis in biology \cite{pattern_form1, pattern_form2}. After the discovery of the origination of pattern formation in RD systems by Alan Turing \cite{turing}, such models have been studied extensively to elucidate biological patterns \cite{bio_pattern1,bio_pattern2,bio_pattern3} in shell, animal pigmentation, fish skin, and chemical processes such as the Belousov-Zhabotinsky reaction \cite{bz}, etc. Pattern formation in ecology \cite{eco_pattern1, eco_pattern2,  eco_pattern3, eco_pattern4, eco_pattern5, eco_pattern6, eco_pattern7, eco_pattern8}, especially in the study of prey-predator systems has received significant attention among the mathematical biologists to unravel the spatiotemporal behavior of interacting species in an ecosystem.

Spatiotemporal patterns emerging in a system of interacting populations can be of either Turing or non-Turing types. Turing patterns, as discussed in the most celebrated paper of Turing \cite{turing}, emerge due to the diffusion driven instability of the homogeneous equilibrium state which results in temporally stable and spatially inhomogeneous solutions of the RD system. On the other hand, spiral wave pattern is one example of the non-Turing patterns that is usually formed near the Hopf bifurcation  boundary and emerges when the spatially uniform wave velocity gets locally disrupted. Spirals are predominantly present in many examples of complex systems and have widespread applicability in spatially extended systems of excitable or self-oscillatory units. They have been experimentally detected in several chemical processes such as Belousov-Zhabotinsky reaction \cite{bz1}, carbon monoxide oxidation on a platinum surface \cite{co}, or in biological processes of aggregating slime molds \cite{slime}, Gierer-Meinhardt model \cite{gm}, etc. Moreover, they are also pertinent to several pathological and neuronal contexts, such as cardiac arrhythmia \cite{cardiac}, human visual cortex \cite{cortex}, hippocampal slices \cite{hippo}, etc. Besides individual analysis of the patterns resulting from Turing or Hopf instabilities, there are also studies involving the simultaneous interaction of these two instabilities. On this note, Rovinsky and Menzinger \cite{turing-hopf} reported that complex spatiotemporal patterns arise due to the nonlinear interaction between Hopf and Turing instabilities. Plenty of researchers explored spatially extended prey-predator systems considering diverse functional responses and reported the impact of Turing and Turing-Hopf instabilities \cite{turing-hopf1, turing-hopf2, turing-hopf3, turing-hopf4, turing-hopf5, turing-hopf6}. 

Furthermore, the emergence patterns of synchronization among the interacting dynamical units in a large complex system has been the focus of research due to its omnipresence in different contexts of biology, ecology, physiology, technology, etc. \cite{synch}. One of the most captivating phenomena that has been reported extensively in the last two decades is the emergence of hybrid chimera patterns consisting of both coherent and incoherent sub-populations. In the coherent region the dynamical variables follow a smooth profile whereas a random distribution of the variables is noticed in the incoherent region.  Primarily observed by Kuramoto and Battogtokh \cite{kuramoto} in a network of complex Ginzburg-Landau systems, later such states have been explored widely in a variety of systems with diverse interaction topologies. Chimera states are found in the presence of local \cite{chimera_local1, chimera_local2, chimera_local3, chimera_local4}, nonlocal \cite{chimera_nonlocal1, chimera_nonlocal2, chimera_nonlocal3, chimera_nonlocal4, chimera_nonlocal5} and global \cite{chimera_global1, chimera_global2, chimera_global3, chimera_global4} interactions among the oscillatory units of phase oscillators, neuron models, limit cycle oscillators and chaotic oscillators. Apart from theoretical studies, chimeras have been detected in several experimental investigations \cite{chimera_exp1, chimera_exp2, chimera_exp3, chimera_exp4, chimera_exp5}. These states are closely related to several pathological brain conditions \cite{epileptic}, like epilepsy, schizophrenia, etc., and are pertinent to the phenomenon of unihemispheric slow-wave sleep \cite{unihemi}, noticed in some aquatic animals and migratory birds. 

Depending on the attributes of the spatiotemporal distribution of the constituting elements associated with a system of interacting identical dynamical units, chimera states are categorized as amplitude chimera \cite{amp_chim1, amp_chim2, amp_chim3, amp_chim4}, traveling chimera \cite{travel_chim1, travel_chim2, travel_chim3}, globally clustered chimera \cite{glob_clust}, breathing chimera \cite{breathing1, breathing2, breathing3}, etc. In this regard, exploration on the emergence of amplitude mediated chimera state by Sethia et al. \cite{amc_prl, amc_pre} is noteworthy. These two articles reported for the first time the observance of amplitude mediated chimera states respectively in nonlocally and globally coupled oscillators. In contrast to classical chimera states, where the coherence-incoherence behaviors emerge depending on the phases of the constituents, in amplitude mediated chimera states \cite{amc1, amc2} both the amplitudes and the phases of the oscillators exhibit random distribution in the incoherent domain. Interestingly, the appearance of chimera states has also been explored in a population exhibiting steady state dynamics, which are termed as chimera death states \cite{amp_chim1, chim_dth1, chim_dth2}. Like chimera state, in chimera death state the group of oscillators is subdivided into coexisting domains of coherent and incoherent steady states. Besides chimera death, the interacting dynamical units can also exhibit the appearance of incoherent oscillation death states when the individual units of the coupled system populate randomly distributed multiple branches of inhomogeneous steady states. Most of the previous findings indicate the occurrence of such diverse collective phenomena in a network of coupled dynamical systems interacting through a nonlocal or a global coupling. However, they are yet to be observed in a spatially extended system coupled through local diffusion.

Again, in two-dimensional spatially extended reaction-diffusion systems nonlocality in the coupling induces spiral wave chimera states possessing phase randomized units in the spiral core enclosed by spiral arms containing phase locked units. Unfolding of such chimera patterns was first reported by Shima and Kuramoto \cite{shima} in a three-component reaction-diffusion model where the presence of nonlocality in the coupling induced the anomalous spiral dynamics. Later, the appearance of such exotic chimera patterns was verified theoretically in a handful \cite{sw1, sw2, sw3, sw4, sw5, sw6} of RD and non-RD systems. Apart from these studies, there exist some experimental evidences \cite{sw7, sw8, sw9} which reveal the emergence of spiral chimera states in coupled Belousov-Zhabotinsky chemical oscillators.

However, all these explorations, whether in RD or non-RD systems, primarily focus on the type of coupling configuration among the interacting units for the development of spiral chimera states. Nonlocality among the coupled dynamical elements is believed to be the conventional way for generating such fascinating patterns. Though later, the authors of ref.\cite{sw_local} reported that the spirals with a phase randomized core can indeed appear in locally coupled reaction-diffusion systems also, but the interacting population must be communicating indirectly through a diffusive environment. In contrast to this, recently in \cite{sw_epjst}, the authors revealed the existence of anomalous spiral chimera structure in a two-component prey-predator RD system in the presence of both self- and cross-diffusion without considering any common diffusive medium. 

In the present article, we uncover the development of multiple collective dynamics as well as the transition from spiral motion to chimeric spiral pattern mediated through the concurrence of Turing and Hopf instabilities of the considered RD system. Particularly, we investigate a two-component prey-predator model in the presence of Holling type III functional response where the species are allowed to diffuse over a two-dimensional landscape. Using linear stability analysis we derive the necessary parametric conditions for the Hopf and Turing bifurcations, which play crucial role in forming patterns in RD systems. We mainly focus on the parameter space where the spatially homogeneous steady states of the population dynamics model become unstable in the presence of both Hopf instability as well as Turing instability. Coherent stationary spirals emanate beyond the Hopf bifurcation boundary in an extensive region of the parametric domain. In aquatic species communities, emergence of spiral patterns can be related with the existence of dipole-like structure in the plankton distribution which is formed due to the rotary motions of the plankton patches in the ocean \cite{plankton_siam}. Spiral waves are also observed in host-parasitoid systems. Besides numerical simulation results, we also succeed to derive the linear amplitude equation in order to describe the formation of spiral patterns and validate the results analytically. The interaction between the oscillatory dynamics appearing due to Hopf bifurcation and the inhomogeneous stationary states formed due to Turing bifurcation combine to result in some complex spatiotemporal patterns which have been reported earlier \cite{turing-hopf1, turing-hopf2, turing-hopf3}. Here we reveal that the simultaneous interplay of both the instabilities for suitable choices of parameter values develops a randomized core inside the coherent spiral arm, resembling the structure of spiral chimera pattern. Such transition mechanism from coherent spiral motion to hybrid spiral chimera structure through the concurrence of Turing and Hopf instabilities in spatially extended RD systems has not been explored so far to the best of our knowledge. Moreover, we confirm that the coexistence of such coherence-incoherence pattern is not only observed in the phases but also in the amplitudes of the constituents, thereby resembling an amplitude mediated chimera state. We verify this pattern by calculating the finite time average of the individual dynamical units which reveals the drift in the average values of the incoherent subpopulation from that of the coherent subpopulation. However, aside from various spiral dynamics we also notice the occurrence of several steady state dynamics, such as  the coexistence of spatially coherent-incoherent oscillation death states termed as chimera death states and the existence of spatially incoherent oscillation death states. In addition, completely synchronized temporal oscillations are also detected over an adequate region of the parameter space. In ecological species dispersal networks, appearance of such multiple dynamical patterns can be related with the phenomenon of species invasion, colonization which can explain persistence and diversity of species in an ecosystem. We estimate the emergence of these diversified dynamics by calculating the strength of the incoherent subpopulations for each and every recognizable pattern in the two-dimensional medium. Spatial arrangement of all such distinguishable patterns along with their spatiotemporal behaviors are numerically demonstrated for a comprehensive parametric range.

The rest of the paper is organized as follows. In Sec. \ref{sec2} we introduce the mathematical model for the chosen prey-predator  system over a two-dimensional diffusive medium. Section \ref{sec3} is devoted to the linear stability analysis of the considered system in the absence or presence of the diffusive terms. 
Section \ref{sec5} includes a detailed discussion on the emergence of diverse collective patterns and their characterization over an extensive parameter region. Finally, in Sec. \ref{sec6} we summarize our results and provide suitable conclusions. In Appendix \ref{appendix} we derive the linear amplitude equation to analytically deduce the spiral wave pattern.

\section{Model description} \label{sec2}
In ecology, the generalized Lotka-Volterra type prey-predator models are used widely to understand the complex interaction behavior in multi-species population dynamics in a better way. For our study, we choose a modified prey-predator population model in the presence of Holling type III functional response. Specifically, here we incorporate one additional term in the predator equation that corresponds to the intra-specific competition between the predators \cite{intra1, intra2}. The intra-predator interaction term helps in reducing the rate of predation by decreasing the predator density and at the same time increasing their mortality rates. The above-mentioned assumptions in the presence of diffusion lead to the formulation of a mathematical model which takes the form of a RD system as given by
\begin{equation}
\begin{array}{lcl}\label{eq1}
\frac{\partial u}{\partial t} = a\Big[u(1-\frac{u}{k}) - \frac{bu^2v}{1+u^2}\Big] + D_u\nabla^2u,\\
\frac{\partial v}{\partial t} = \frac{cu^2v}{1+u^2} - dv - \eta v^2 + D_v\nabla^2v.
\end{array}
\end{equation}

Here $u(x,y,t)$ and $v(x,y,t)$ correspond to the prey and predator population densities, respectively, at position $(x,y)$ and time $t$, on a bounded domain $\mathfrak{D} \in \mathbb{R}^2$ with closed boundary $\partial \mathfrak{D}$. The prey species follows a logistic growth with carrying capacity $k$. The consumption of prey by the predator is represented by a Holling  III functional response $\frac{u^2v}{1+u^2}$. Here, $b$ is the maximum per capita predation rate and  $a$ is a multiplicative factor associated with the prey growth equation. The scalar $c$ denotes the fraction of prey's biomass converted into predator's biomass via predation, $d$ is the natural mortality rate of the predator, and $\eta$ corresponds to the intra-predator interaction. The diffusion of both the species over the two-dimensional domain $\mathfrak{D}$ is represented by the Laplacian operator $\nabla^2$, where $D_u, D_v$, respectively signify the diffusivity coefficients of prey and predator populations. We analyze the model under the assumption of non-negative initial conditions along with zero-flux boundary conditions
\begin{equation}
\begin{array}{lcl}\label{eq2}
u(x,y,0) > 0, v(x,y,0) > 0, ~~~(x,y) \in \mathfrak{D} \\
\frac{\partial u}{\partial \hat{n}} = \frac{\partial v}{\partial \hat{n}} = 0, ~~~(x,y)\in \partial \mathfrak{D},
\end{array}
\end{equation}
where $\hat{n}$ is the outward unit normal vector of the boundary $\partial \mathfrak{D}$. For numerical simulations, we discretize the domain on a $N\times N$ two-dimensional lattice with $N = 249$ lattice points in each direction. In order to solve Eq. \eqref{eq1} numerically, we adopt the predictor-corrector method \cite{predict-correct} with alternating directions approach considering spatial step $h = 0.25$ and integration time step $dt = 0.025$. Other appropriate choices of $N, h$ and $dt$ could also give rise to the desired dynamics.

\section{Linear stability analysis} \label{sec3}
In the following subsections, we will analyze the equilibrium points and the stability of the temporal (i.e., $D_u = D_v = 0$) as well as the spatiotemporal (i.e., $D_u \neq 0, D_v \neq 0$) systems.

\subsection{Temporal model}\label{temporal}

\begin{figure}[ht]
	\centerline{
		\includegraphics[scale=0.45]{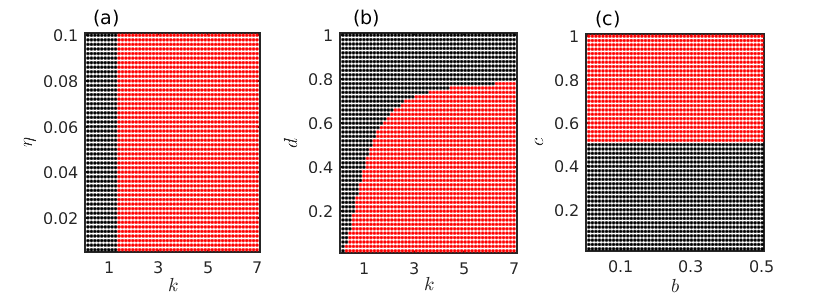}}
	\caption{The number of feasible spatially homogeneous coexistence equilibrium points of the system \eqref{eq1} in (a) $k-\eta$, (b) $k-d$, and (c) $b-c$ parameter spaces. The parameter values are fixed at $k=6.5, b = 0.1923, c = 0.8, d = 0.5$ and $\eta = 0.018$ when any two of them are varied. Red and black colors correspond to the existence of only one and zero number of coexistence solution, respectively.}
	\label{fig1}
\end{figure}

The equilibrium point $(u_*,v_*)$ of the system \eqref{eq1} in the absence of diffusion (i.e., $D_u = D_v = 0$) is given by the solution of the equations $f(u_*,v_*) = 0$ and $g(u_*,v_*) = 0$, where 
\begin{equation}
\begin{array}{lcl}\label{eq3}
f(u,v) = u(1-\frac{u}{k}) - \frac{bu^2v}{1+u^2},~~~
g(u,v) = \frac{cu^2v}{1+u^2} - dv - \eta v^2.
\end{array}
\end{equation} 
Solving Eq. \eqref{eq3} we can easily detect the following different possible equilibrium points of the system:\\
(i) Extinction of both the prey and predator population: $E_{00}=(0,0)$.\\
(ii) Existence of only prey population: $E_{u0} = (k,0)$, where the prey population varies depending on the carrying capacity $k$.\\
(iii) Coexistence of both the populations: $E_{uv} = (u_s,v_s)$, where $v_s = \frac{1}{\eta}\Big(\frac{cu_s^2}{1+u_s^2}-d\Big)$ and $u_s$ is a positive root of the following quintic equation:
\begin{equation}\label{eq4}
u_s^5 - ku_s^4 + \Big(2+\frac{kb(c-d)}{\eta}\Big)u_s^3 - 2ku_s^2 + \Big(1-\frac{kbd}{\eta}\Big)u_s - k = 0.
\end{equation}
Based on the parametric constraints, the number of interior equilibrium points (coexistence of both the prey and predator populations) can vary from zero to five. Here we plot the number of feasible co-existence equilibrium points $(u_s, v_s)$ in different parameter spaces as depicted in Fig. \ref{fig1}. The following parameter values $k=6.5, b = 0.1923, c = 0.8, d = 0.5$ \cite{epjst_connect} are fixed in such a manner that the system (\ref{eq1}) exhibits oscillatory dynamics in the absence of diffusion with proper choice of the parameters $a$ and $\eta$. Looking into the $k-\eta$ parameter space plotted in Fig. \ref{fig1}(a), we fix the value of intra-predator interaction rate $\eta$ at $\eta = 0.018$ for the remaining numerical simulations. In the figure, the red region corresponds to the existence of only one interior equilibrium point while the black region signifies the absence of coexistence solution (i.e., negative real roots of Eq. \eqref{eq4}). Thus the parameter spaces depicted in all the figures in Fig. \ref{fig1} indicate that the system \eqref{eq1} has maximally one interior equilibrium point in the absence of diffusion for our chosen values of the parameters.

The eigenvalue $\lambda$ of the Jacobian matrix $J$ of the linearized system determining the stability \cite{book_ml} of the coexistence equilibrium $(u_s,v_s)$, satisfies the equation $\lambda^2 - Tr(J)\lambda + \Delta_J = 0$. Thus,
\begin{equation}\label{eq5}
\lambda_{1,2} = \frac{Tr(J) \pm \sqrt{Tr(J)^2-4\Delta_J}}{2},
\end{equation}
where $Tr(J)$ and $\Delta_J$ are the trace and determinant of the Jacobian matrix given by 
\begin{equation}\label{eq6}
Tr(J) = af_u+g_v, ~~~\Delta_J = a(f_ug_v-f_vg_u).
\end{equation}
In Eq. \eqref{eq6} $f_u, f_v, g_u, g_v$ are the partial derivatives of the functions $f, g$ with respect to $u, v$, respectively. The equilibrium point is stable if  $Tr(J)$ and $\Delta_J$ evaluated at $(u_s,v_s)$ necessarily satisfy the conditions $Tr(J) < 0$, $\Delta_J >0$, and unstable otherwise. The system exhibits a Hopf bifurcation when the eigenvalues cross the imaginary axis at a point where $Tr(J) = 0$, and the threshold for the bifurcation is derived as $a_0 = -\frac{g_v}{f_u}$.

\subsection{Spatiotemporal model}

Now we investigate the condition for Turing instability of the spatiotemporal system \eqref{eq1} where the spatially homogeneous stable steady state $(u_s,v_s)$ becomes unstable to small amplitude perturbations in the presence of diffusion coefficients $D_u$ and $D_v$.
Linearizing the system around the interior equilibrium point $(u_s, v_s)$, we can write
\begin{equation}
\begin{array}{lcl}\label{eq7}
\frac{\partial \tilde{u}}{\partial t} = af_u \tilde{u} + af_v \tilde{v} + D_u \nabla^2 \tilde{u},\\\\
\frac{\partial \tilde{v}}{\partial t} = g_u \tilde{u} + g_v \tilde{v} + D_v \nabla^2 \tilde{v},
\end{array}
\end{equation}
where $(\tilde{u},\tilde{v})$ is the small perturbation around $(u_s,v_s)$ such that $u = u_s + \tilde{u}$ and $v = v_s + \tilde{v}$.
We consider, 
\begin{equation}\label{eq8}
\begin{pmatrix}
\tilde{u} \\ \tilde{v}
\end{pmatrix} = \begin{pmatrix}
\tilde{u}_\sigma \\ \tilde{v}_\sigma
\end{pmatrix} e^{\lambda t + i \vec{\sigma}.\vec{r}},
\end{equation}
as the solution of the linearized system $\eqref{eq7}$. Here, $\lambda$ determines the rate of evolution of the perturbation over time $t$, $\vec{\sigma} = (\sigma_x,\sigma_y)$ is the wave number such that $\vec{\sigma}.\vec{\sigma} = \sigma^2$ and $\vec{r} = (x,y)$ is the spatial position in two dimensions. In \eqref{eq8}, $\tilde{u}_\sigma, \tilde{v}_\sigma$ are the amplitudes of the perturbation. The Jacobian $J_D$ of the linearized system \eqref{eq7} is therefore given by
\begin{equation}
J_D = \begin{pmatrix}\label{eq9}
af_u-D_u\sigma^2 && af_v \\ g_u && g_v-D_v\sigma^2
\end{pmatrix}.
\end{equation}
Hence, the characteristic equation can be written as 
\begin{equation}\label{eq10}
\lambda^2 - Tr(J_D) + \Delta_{J_D} = 0,
\end{equation}
where $Tr(J_D) = Tr(J) - (D_u+D_v)\sigma^2$ and $\Delta_{J_D} = D_uD_v\sigma^4 - (D_ug_v+aD_vf_u)\sigma^2 + \Delta_J$.
For the emergence of Turing instability, $(u_s,v_s)$ must be stable in the absence of diffusion which guarantees that $Tr(J) < 0$ and $\Delta_J > 0$. Therefore, from the expression of $Tr(J_D)$, it is clear that $Tr(J_D) < 0$. Obviously, in order to induce instability in the spatiotemporal model \eqref{eq1} in the presence of diffusion coefficients $D_u$ and $D_v$, $\Delta_{J_D}$ must be negative, i.e., $D_uD_v\sigma^4 - (D_ug_v+aD_vf_u)\sigma^2 + \Delta_J < 0$. But, the diffusion coefficients $D_u, D_v$ as well as $\Delta_J$ all are positive, so necessarily the coefficient of $\sigma^2$ in the expression of $\Delta_{J_D}$ should be positive to generate Turing instability. Thus, 
\begin{equation}\label{eq11}
(D_ug_v+aD_vf_u) > 0
\end{equation}
is the necessary condition for the instability of the equilibrium point $(u_s,v_s)$ in the presence of diffusion. Combining the condition \eqref{eq11} and $Tr(J) < 0$, it eventually implies that $D_v > D_u$ (since $f_u(u_s,v_s) > 0$ and $g_v(u_s,v_s) < 0$) for the equilibrium to be unstable.

\begin{figure}[ht]
	\centerline{
		\includegraphics[scale=0.45]{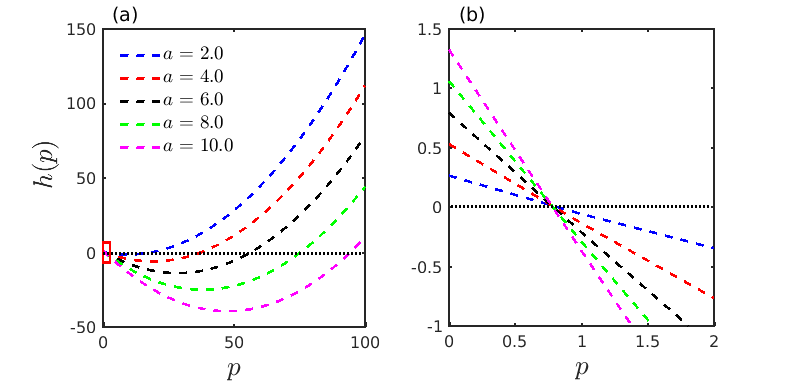}}
	\caption{Plot of $h(p)$ with respect to $p$ for different values of the system parameter $a$. (a) As $a$ increases the width of the interval for which $h(p) < 0$ increases, (b) zoomed version of the red rectangular region in (a). Here the black dotted line corresponds to $h(p) = 0$. The parameter values considered here are $k=6.5, b = 0.1923, c = 0.8, d = 0.5$, $\eta = 0.018, D_u = 0.01$ and $D_v = 1.8$.}
	\label{fig2}
\end{figure}

Now, we consider the expression 
\begin{equation}\label{eq12}
h(p) = D_uD_vp^2 - (D_ug_v+aD_vf_u)p + \Delta_J,
\end{equation}
which is a quadratic equation in $p$ with $p = \sigma^2$. Here we plot the quadratic expression $h(p)$ with respect to $p$ in Fig. \ref{fig2}(a) for different values of the parameter $a$. It is easily discernible from the figure that as the value of $a$ increases the curve of $h(p)$ moves downwards and thereby the width of the interval for which $h(p) < 0$ increases. So, it is evident that the possibility of occurrence of Turing instability increases as the system parameter $a$ is increased. For better visibility in Fig. \ref{fig2}(b), we show the zoomed version of the red rectangular region near $p=1$ depicted in Fig. \ref{fig2}(a). This figure confirms that the determinant $h(0)$ of the Jacobian of the non-spatial system is positive for all considered values of $a$. The graph of $h(p)$ with respect to $p$ represents a parabolic structure for which the minimum is reached at $p = p_{m}$, following the relation $\frac{dh(p)}{dp} = 0$. Solving this we obtain the threshold of $p$ as $p_{m} = \frac{D_ug_v+aD_vf_u}{2D_uD_v}$ at which the function $h(p)$ attains its minima, as $\frac{d^2h(p)}{dp^2} > 0$, and the minimum value is given by 
\begin{equation}\label{eq13}
h_{min}(p_m) = -D_uD_vp_{m}^2 + \Delta_J.
\end{equation}
Using the condition \eqref{eq13} we can derive the expression for the critical value of diffusion coefficient $D_v$ beyond which Turing instability emerges in the spatiotemporal system and this value is obtained as
\begin{equation}\label{eq14}
D_v^{Crit} = \frac{D_u}{af_u^2}\Big[ (f_ug_v-2f_vg_u) \pm 2\sqrt{f_vg_u(f_vg_u-f_ug_v)}\Big].
\end{equation}

\begin{figure}[ht]
	\centerline{
		\includegraphics[scale=0.5]{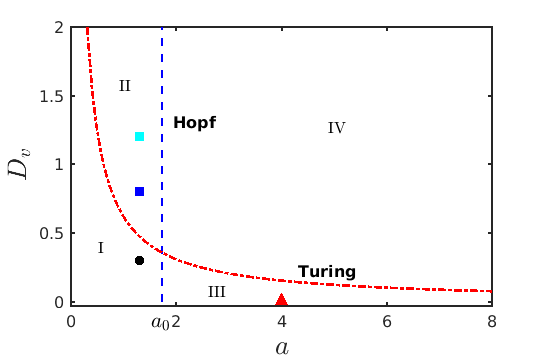}}
	\caption{Critical curves for the Hopf and Turing bifurcations in $a-D_v$ parameter space of the system \eqref{eq1}. Blue dashed vertical line corresponds to the Hopf bifurcation threshold $a_0$ and the red curve is associated with the Turing condition obtained in \eqref{eq14}. Region I: system reaches the spatially homogeneous stationary state $(u_s,v_s)$; Region II: spatially homogeneous steady state loses its stability and the system reaches inhomogeneous steady states confirming the appearance of Turing patterns; Region III: Hopf instability occurs and the system exhibits limit cycle oscillations; Region IV: both Hopf and Turing instabilities occur. Other parameter values are same as mentioned in Fig. \ref{fig2}. Markers plotted in the figure correspond to the parameter values used in Fig. \ref{fig4}.}
	\label{fig3}
\end{figure}
\begin{figure}[ht]
	\centerline{
		\includegraphics[scale=0.48]{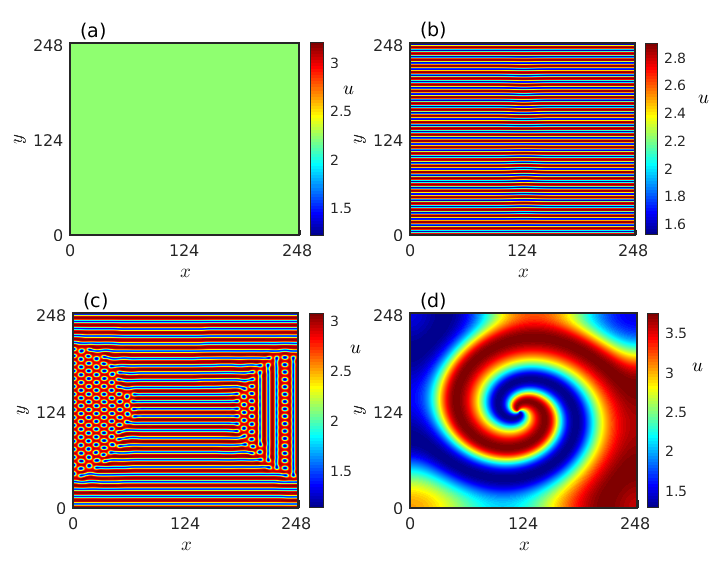}}
	\caption{Spatial patterns of the prey species observed in the regions I, II and III of Fig. \ref{fig3} for which the parameter values are fixed at (a) $a=1.3, D_u = 0.01, D_v = 0.3$ (black circle in Fig. \ref{fig3}): spatially homogeneous stable stationary state, (b) $a=1.3, D_u = 0.01, D_v = 0.8$ (blue square in Fig. \ref{fig3}): stripes pattern, (c) $a=1.3, D_u = 0.01, D_v = 1.2$ (cyan square in Fig. \ref{fig3}): combination of stripes and spots, and (d) $a=4.0, D_u = 0.01, D_v = 0.01$ (red triangle in Fig. \ref{fig3}): spiral pattern. Other parameter values are same as mentioned earlier in the text. The color bars represent the prey population density $u$.}
	\label{fig4}
\end{figure}

Figure \ref{fig3} depicts the critical value \eqref{eq14} in red for a wide range of the parameter values of $a$. Here the blue dashed vertical line corresponds to the Hopf bifurcation threshold $a_0$. These two Hopf and Turing instability curves intersect at a particular point and divide the whole parametric space into four distinct domains. Below the Turing curve in the domain I, the system reaches the spatially homogeneous stationary state $(u_s,v_s)$, while in domain II above the Turing curve the spatially homogeneous steady state loses its stability and the system reaches inhomogeneous steady states which confirms the appearance of Turing patterns. On the other side, beyond the Hopf boundary in domain III, Hopf instability occurs and the system exhibits limit cycle oscillations, whereas the most interesting non-trivial dynamics is observed in domain IV where both Hopf and Turing instabilities take place. Figure \ref{fig4} depicts the spatial patterns of the prey species for some exemplary parameter values from the regions I, II and III. To generate the spatial patterns, the initial distribution of the species is considered as
\begin{equation}
\begin{array}{lcl}
u(x,y,0) = u_s - \epsilon_1(y - 0.5(N-1)),\\
v(x,y,0) = v_s - \epsilon_2(x - 0.5(N-1)),
\end{array}
\end{equation}
where $(x,y)$ is considered as the grid index and $\epsilon_1, \epsilon_2$ are small fluctuations added to the homogeneous steady state. We take $\epsilon_1 =  \epsilon_2 = 10^{-4}$. The appearance of spatially homogeneous steady state in region I is portrayed in Fig. \ref{fig4}(a) for the parameter value $a=1.3$ and diffusion strengths $D_u = 0.01, D_v = 0.3$. Figures \ref{fig4}(b) and \ref{fig4}(c) correspond to the Turing patterns when the parameter values are chosen from region II. Stripe pattern is depicted in Fig. \ref{fig4}(b) for the parameter values $a=1.3, D_u = 0.01$ and $D_v = 0.8$ while a combination of stripes and spots is detected for the parameter values $a=1.3, D_u = 0.01$ and $D_v = 1.2$ in Fig. \ref{fig4}(c). An intriguing spiral pattern is illustrated in Fig. \ref{fig4}(d) for the parameter values $a=4.0, D_u = 0.01, D_v = 0.01$ set in region III. In the following section, we will explore the rich dynamical patterns observed on the right of the Hopf boundary below and above the Turing curve.

\section{Diverse collective states beyond Hopf boundary} \label{sec5}
In this section, we explore the emergence of various spatiotemporal patterns and their dynamical characteristics in a broad region of the parameter space spanned by the diffusion coefficients. 

\subsection{Emergence of spatial patterns for $a = 2.0$}
\begin{figure}[ht]
	\centerline{
		\includegraphics[scale=0.35]{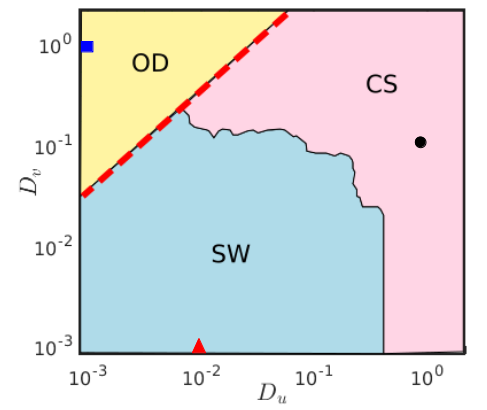}}
	\caption{Emergence of different spatial patterns in $D_u-D_v$ parameter space for parameter value $a=2.0$. The regions shaded with sky blue, pink and yellow colors correspond to the regions of spiral wave (SW) oscillations, completely synchronized (CS) oscillations and inhomogeneous steady states (OD), respectively. Red dashed line represents the critical boundary above which Turing instability persists in the system. Spatiotemporal evolution of the system corresponding to the parameter values marked with triangle, circle and square are depicted in Fig. \ref{fig6}.}
	\label{fig5}
\end{figure}

From a detailed numerical analysis, we find that three different spatial patterns arise in the $D_u-D_v$ parameter space near the Hopf boundary for the parameter value $a=2.0$ as presented in Fig. \ref{fig5}. The regime diagram is obtained by looking into the snapshot at the corresponding diffusion parameter values at the final time when the system is simulated up to $t = 5 \times 10^4$ time units. In addition, we also checked the temporal as well as the spatiotemporal dynamics in the time interval $[4.99 \times 10^4, 5 \times 10^4]$ to classify the states as oscillatory or non-oscillatory. The red dashed line corresponds to the critical threshold $D_v^{Crit}$ for Turing instability as obtained from Eq. \eqref{eq14}. The region below the critical line corresponds to the parameter values for which oscillatory dynamics is present, whereas in the region above the critical line steady state dynamics persists. For comparatively smaller values of diffusion coefficients $D_u, D_v$, spiral wave pattern is noticed. The corresponding parameter region is shaded with sky blue color. Higher values of prey and predator diffusion strengths (the region shaded in pink) below the red dashed line lead to a pattern where completely synchronized temporal oscillations are obtained. The region shaded in yellow corresponds to the Turing instability region where the system settles down to inhomogeneous steady state dynamics confirming the appearance of various Turing patterns. From the figure it is evident that the analytically derived boundary (marked with red dashed line) partitioning the oscillatory and the non-oscillatory regimes is in perfect accordance with the numerically simulated parameter space. 

\begin{figure}[ht]
	\centerline{
		\includegraphics[scale=0.43]{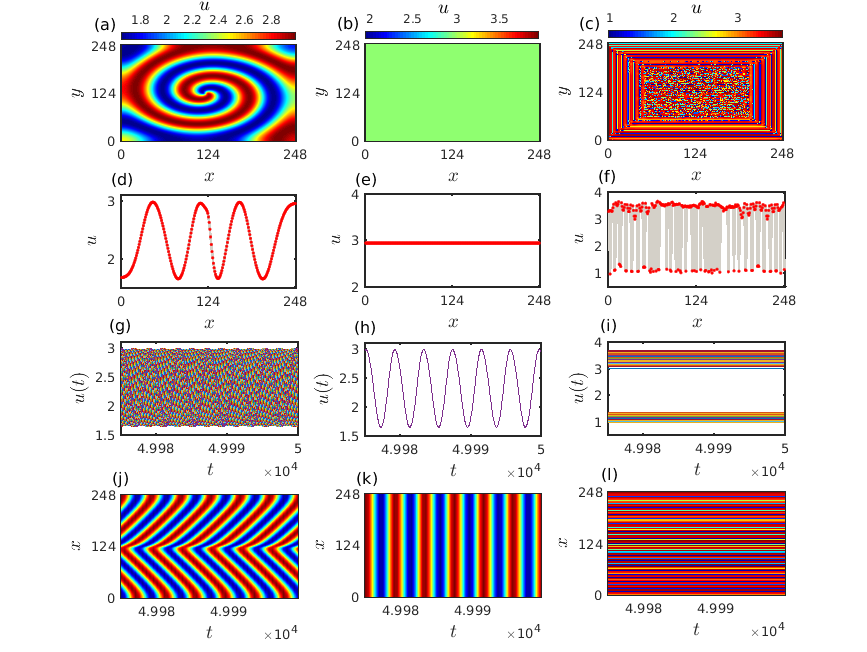}}
	\caption{Evolution of various spatial patterns corresponding to the markers in Fig. \ref{fig5} for prey growth rate $a=2.0$. (a,b,c) Snapshots of the prey species in the two-dimensional diffusive medium at a particular time instant $t= 5 \times 10^4$, (d,e,f) horizontal cross-section along the center is plotted in red dots with respect to the number of discretized grids in each direction, (g,h,i) temporal dynamics of prey species of all the $249$ sites along the cross-section corresponding to each pattern, (j,k,l) spatiotemporal evolution of the corresponding prey densities. Diffusion coefficients are fixed at $D_u = 0.01, D_v = 0.001$ (left panel, corresponding to red triangle in Fig. \ref{fig5}), $D_u = 1.0, D_v = 0.1$ (middle panel, corresponding to black circle in Fig. \ref{fig5}), and $D_u = 0.001, D_v = 1.0$ (right panel, corresponding to blue square in Fig. \ref{fig5}). The associated colorbars quantify the prey densities $u$ at that instant of time. The color codings for the bottom row figures are same as that of the corresponding upper row figures.} 
	\label{fig6}
\end{figure}

In the following Fig. \ref{fig6}, we demonstrate the spatial snapshots of the various patterns as mentioned in Fig. \ref{fig5} and their evolution over time. Left, middle and right columns correspond to the evolution of the prey species exhibiting spiral, synchronized oscillatory and inhomogeneous steady state dynamics, respectively. Snapshots of these patterns in the prey population over a two-dimensional diffusive medium are presented in Figs. \ref{fig6}(a,b,c) at a particular instant of time. The corresponding colorbars quantify the densities of the prey population $u$ at that time instant. To better understand the organization among the species for each pattern we have taken a horizontal cross-section along the central line and plotted them in Figs. \ref{fig6}(d,e,f) with respect to the number of discretized grids in each direction. Figures \ref{fig6}(g,h,i) in the third row exhibit the temporal dynamics of the prey species along the horizontal cross section and finally, Figs. \ref{fig6}(j,k,l) in the bottom row correspond to the spatiotemporal evolution of the patterns among the prey species along the cross-section. Existence of stable spiral pattern is depicted in Fig. \ref{fig6}(a) for diffusion coefficients $D_u = 0.01$ and $D_v = 0.001$. The frequency of rotation of the spiral wave decreases and the wavelength of the spiral increases when either one or both the diffusion strengths are increased (figure not shown here). The one dimensional cross-section of the spiral pattern presented in Fig. \ref{fig6}(d) reflects a coherent oscillatory structure among the species in the discrete grids. Temporal dynamics in the form of limit cycle oscillations of these species along the cross section is clearly visible in Fig. \ref{fig6}(g). The observed spiral wave starts to propagate from the core and expands in the outward direction which can be perceived from the spatiotemporal evolution of the prey species along the cross-section depicted in Fig. \ref{fig6}(j). Figure \ref{fig6}(b) reveals the uniform spatial density of the prey species in the two-dimensional medium for diffusion strengths $D_u = 1.0$ and $D_v = 0.1$ and their corresponding horizontal cross-section along the center line is presented in Fig. \ref{fig6}(e). The temporal as well as the spatiotemporal evolutions depicted in Figs. \ref{fig6}(h) and \ref{fig6}(k) confirm the existence of completely synchronized limit cycle oscillations with uniform spatial density of all the species. These two figures clearly depict that at any particular instant of time the spatial densities of all the species remain constant. That is why the spatial snapshot in Fig. \ref{fig6}(b) takes only a single value corresponding to the light green color from the colorbar. Emergence of inhomogeneous steady states in the two-dimensional diffusive environment is demonstrated in Fig. \ref{fig6}(c) resulting from the Turing instability at the diffusion coefficient values $D_u = 0.001$ and $D_v = 1.0$. Horizontal cross-section of the prey species $u$ along the center line presented in Fig. \ref{fig6}(f) signifies that the species density keeps fluctuating between the upper and the lower branches while maintaining a steady state temporal dynamics. Two distinguishable clusters of prey densities are clearly visible from the temporal dynamics presented in Fig. \ref{fig6}(i). Such kind of temporal pattern where the inhomogeneous steady states are subdivided into more than one distinguishable clusters are known as multicluster oscillation death states \cite{mcod}. The corresponding spatiotemporal evolution of these multicluster oscillation death states is portrayed in Fig. \ref{fig6}(l).

\subsection{Emergence of amplitude mediated spiral chimera}

\begin{figure}[ht]
	\centerline{
		\includegraphics[scale=0.3]{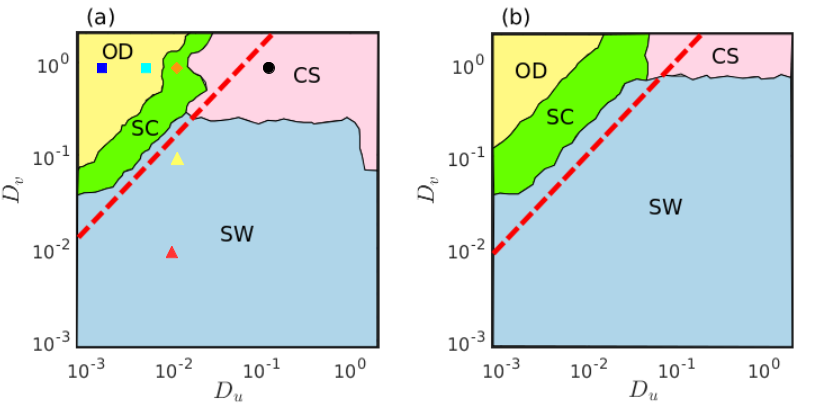}}
	\caption{Emergence of diverse spatial patterns in $D_u-D_v$ parameter space for prey growth rate (a) $a=4.0$, (b) $a=6.0$. Regions shaded with sky blue, green, pink and yellow colors respectively correspond to the regions of spiral wave oscillations, spiral chimera, completely synchronized oscillations and inhomogeneous steady states. Red dashed line corresponds to the analytically derived critical boundary above which inhomogeneous steady state dynamics is expected in the system. Markers plotted in Fig. \ref{fig7}(a) correspond to the parameter values used in Figs. \ref{fig8}, \ref{fig9} and \ref{fig10}.}
	\label{fig7}
\end{figure}

Besides the above mentioned three distinct categories of spatial patterns, another fascinating spatial organization among the dynamical elements is noticed as the parameter $a$ is shifted away from the Hopf bifurcation boundary. For appropriate choices of prey and predator diffusion strengths $D_u$ and $D_v$, the core of the spiral motion in the two-dimensional space starts to expand. The dynamical units inside the spiral core exhibit anomalous behavior which differentiates the core region from that of the spiral arm where the constituents follow a smooth motion. Simultaneous occurrence of coherent dynamics in the spiral arm and incoherent dynamics in the spiral core discriminates the spatial pattern from that of the ordinary spiral dynamics. Such coexistence of coherent and incoherent behaviors within a spiral structure develops the captivating spiral chimera pattern. Emergence of spiral chimera patterns in $D_u-D_v$ parameter space is manifested in Fig. \ref{fig7} through the parametric region shaded with green color. Other parameter regions shaded with sky blue, pink and yellow colors correspond to similar spatial dynamics as discussed earlier in Fig. \ref{fig5}. Apart from the diffusion strengths associated with the yellow shaded region all other diffusion strengths correspond to oscillatory dynamics of the considered reaction-diffusion system. Two different parameter spaces are depicted in Figs. \ref{fig7}(a) and \ref{fig7}(b), respectively, for two different values of the multiplicative factor $a=4.0$ and $a=6.0$. As mentioned already, the red dashed line delimits the analytically obtained bound $D_v^{Crit}$ for Turing instability. Although, contrary to the case of Fig. \ref{fig5} where the red boundary perfectly matches with the numerically simulated parameter region, a discrepancy is noticed in Fig. \ref{fig7} between the analytical and numerical bounds separating the oscillatory and non-oscillatory regimes. From the figure it is visible that the analytically derived red boundary fails to determine the critical threshold beyond which oscillatory dynamics vanishes and steady state dynamics (yellow shaded region) begins. Such inconsistency is triggered by the simultaneous occurrence of both the Hopf and the Turing instabilities for the considered parameter values of $a$. In the right hand side of the blue vertical line displayed in Fig. \ref{fig3}, whenever $a$ remains very close to the critical value $a_0$, the emergence of Turing instability overrules the Hopf instability, i.e., the oscillatory dynamics is suppressed and the steady state dynamics arises as soon as the diffusion coefficients satisfy the Turing instability condition. That is why for $a=2.0$, once the diffusion coefficients reach the Turing instability condition (Eq. \eqref{eq14}), steady state dynamics replaces the oscillatory dynamics which is reflected perfectly in the parameter space of Fig. \ref{fig5}. However, when $a$ moves away further from $a_0$, the appearance of oscillations due to Hopf bifurcation reveals dominating characteristic and as a result, Eq. \eqref{eq14} fails to  predict the emergence of steady state dynamics of the system \eqref{eq1}. This scenario is demonstrated through the parameter spaces in Fig. \ref{fig7} as there is appreciable difference between the numerically obtained parameter values (yellow region) and the analytically derived bounds for the appearance of steady state dynamics.

\begin{figure}[ht]
	\centerline{
		\includegraphics[scale=0.42]{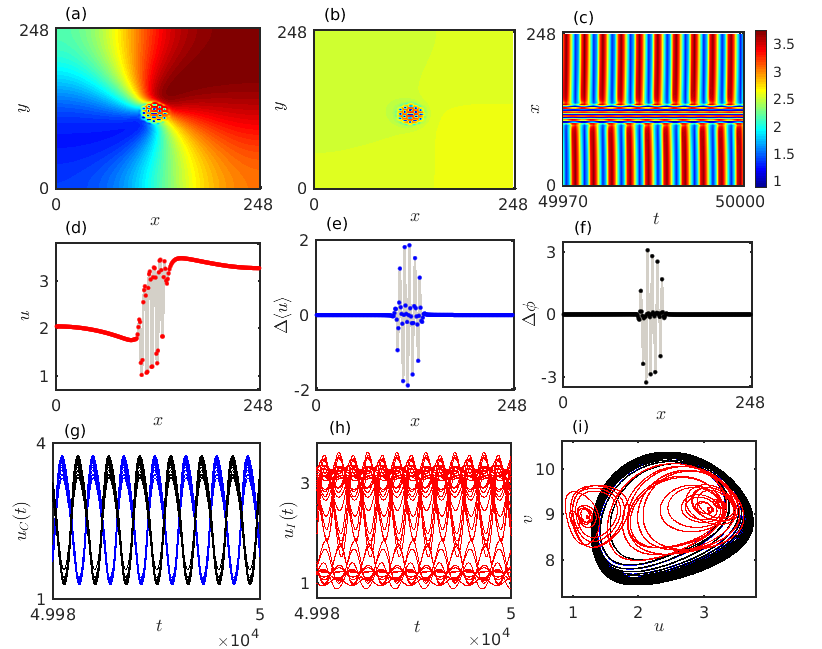}}
	\caption{Evolution of amplitude mediated spiral chimera pattern: (a) snapshot of the prey density $u$ in the two-dimensional space at a particular time instant $t = 5\times 10^4$ depicting incoherence in the spiral core, (b) corresponding finite-time average $\langle u \rangle$ over $T =10^3$ time iterations, (c) spatiotemporal behavior of the species density along the horizontal cross-section through the line $y=124$, (d) snapshot along the horizontal cross-section for a better visualization of the incoherence dynamics around the center, (e) amplitude difference $\Delta \langle u \rangle$ between two adjacent species densities along the horizontal cross-section, (f) corresponding phase difference $\Delta\phi$ between two adjacent constituents along the cross-sectioned profile, (g) temporal dynamics of the coherent subpopulations along the line $y=124$, (h) temporal dynamics of the incoherent subpopulation along the line $y=124$, (i) attractors corresponding to the amplitude mediated chimera state along the cross-section through $y=124$. Prey and predator diffusion strengths are fixed at $D_u=0.01$ and $D_v=1.0$ corresponding to the orange diamond marker plotted in Fig. \ref{fig7}(a).}
	\label{fig8}
\end{figure}

The interplay between the two different instabilities where the oscillations caused by the Hopf bifurcation interact with the inhomogeneous steady states arising due to the Turing bifurcation leads to a fascinatingly complex spiral chimera pattern. We carry out a detailed numerical simulation to analyze the characteristics of such an exotic chimera structure as presented in Fig. \ref{fig8}. Figure \ref{fig8}(a) depicts the typical snapshot of the spiral chimera pattern at a particular time instant $t=5 \times 10^4$ for $a=4.0$, when the diffusion coefficients are fixed at $D_u = 0.01$ and $D_v = 1.0$. For a better understanding of the correlation among the species we take a horizontal cross-section along the center line and plot them in Fig. \ref{fig8}(d). Anomalous species densities around the spiral core surrounded by coherent densities of the prey species is detectable from both the Figs. \ref{fig8}(a) and \ref{fig8}(d). To get further insight about the amplitude of the oscillatory prey densities over the two-dimensional medium, we calculate the finite-time average of the prey densities $u(x,y,t)$ at each grid index $(x,y)$ over $T = 10^3$ time iterations using the formula $\langle u(x,y,t) \rangle = \frac{1}{T} \int_{0}^T u(x,y,t)dt$. We plot the estimated average prey densities $\langle u \rangle$ in the two-dimensional $x-y$ plane in Fig. \ref{fig8}(b) which depicts that apart from the species situated in the spiral core all the remaining prey species oscillate with a nearly identical amplitude, whereas random fluctuations are noticed around the core region. For a clear visualization here also we take the horizontal cross-section along the center line and plot the amplitude differences between two adjacent species densities in Fig. \ref{fig8}(e). Zero values of the amplitude difference $\Delta \langle u \rangle$ suggest coherent amplitude of the prey species densities outside the spiral core while non-zero values of $\Delta \langle u \rangle$ around the core corresponds to incoherent species amplitude. The corresponding spatiotemporal evolution of the species densities along the horizontal cross-section is presented in Fig. \ref{fig8}(c) which reveals that the observed chimera pattern is stationary with respect to time. We further calculate the phases of each of the individual components available in the two-dimensional diffusive medium using the formula $\phi(x,y) = \tan^{-1} \Big(\frac{v(x,y) - v_s}{u(x,y)-u_s}\Big)$. Figure \ref{fig8}(f) delineates the phase differences $\Delta \phi$ between two adjacent units situated along the cross-sectioned profile in Fig. \ref{fig8}(d). Again a zero value of $\Delta \phi$ characterizes coherent subpopulation whereas a non-zero value of the phase difference confirms the presence of incoherent dynamics. Thus the observed chimera pattern exhibits coherence-incoherence dynamics in both the phases and the amplitudes of the constituent elements, and hence resembles the structure of an amplitude mediated chimera state. Besides, we also explore the temporal dynamics of each of the individual elements along the cross-sectioned profile. From the cross-sectioned prey densities it is evident that the coherent subpopulations are separated into two disjoint clusters by the incoherent subpopulations. The time series for the coherent subpopulations are presented in Fig. \ref{fig8}(g), where the two distinct clusters follow two distinguishabe temporal patterns marked with blue and black curves. The figure indicates that the two disjoint clusters of coherent subpopulations are in opposite phases with each other. However, the randomness in the temporal dynamics of the incoherent subpopulations is presented by red curves in Fig. \ref{fig8}(h). The corresponding phase space dynamics of the amplitude mediated chimera state along the cross-section is portrayed in Fig. \ref{fig8}(i). Positive correlation among the coherent subpopulations are prominent from the attractors plotted with blue and black curves, while randomness is detectable among the incoherent group through the plotted attractors in red.

\subsection{Multiple oscillatory and non-oscillatory dynamics for $a=4.0$}

\begin{figure}[ht]
	\centerline{
		\includegraphics[scale=0.45]{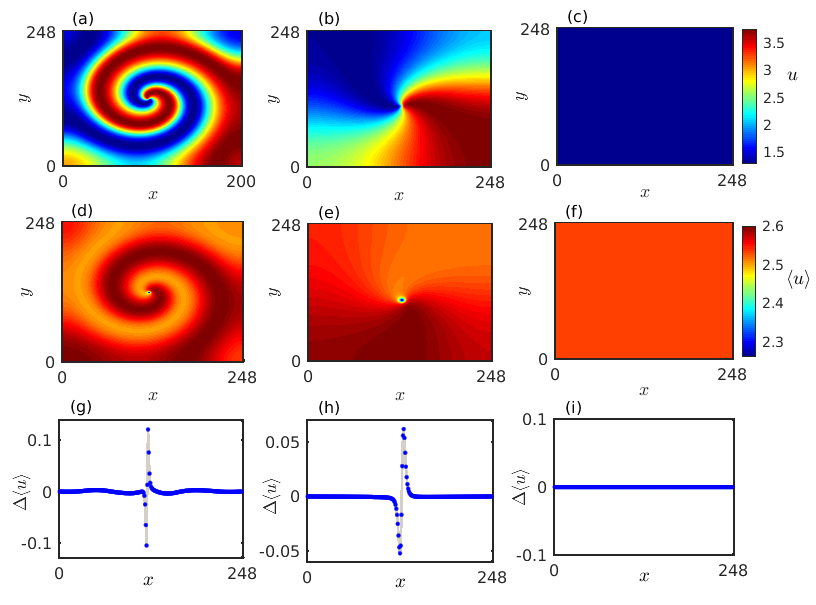}}
	\caption{(a,b,c) Snapshots of temporally oscillating prey densities $u$, displaying various spatial patterns for different combinations of diffusion strengths at a particular time instant $t=5\times 10^4$, (d,e,f) corresponding finite-time average $\langle u \rangle$ of the prey densities associated with each individual pattern over $T=10^3$ time iterations, (g,h,i) amplitude difference $\Delta \langle u \rangle$ between two adjacent species densities along the horizontal cross-section through the line $y=124$. Diffusion coefficients are fixed at values corresponding to the markers depicted in Fig. \ref{fig7}(a): (a,d,g) $D_u = D_v = 0.01$ (red triangle), (b,e,h) $D_u = 0.01, D_v = 0.1$ (yellow triangle), (c,f,i) $D_u = 0.1, D_v = 1.0$ (black circle). }
	\label{fig9}
\end{figure}

Further, we also scrutinize the development of other oscillatory patterns for $a=4.0$ in terms of their average amplitude as illustrated in Fig. \ref{fig9}. The upper panel depicts the typical snapshots of distinct spatial patterns that are oscillating in time for different combinations of the diffusion strengths at a particular time instant $t=5 \times 10^4$. The finite-time averages of each of these individual patterns over $T=10^3$ time iterations are presented in the middle row, while the amplitude differences between the adjacent nodes along the horizontal cross-sections through the center lines are presented in the bottom row. Snapshot of an ordinary spiral is portrayed in Fig. \ref{fig9}(a) for lower values of both the diffusion coefficients $D_u = D_v = 0.01$. Figure \ref{fig9}(d) shows the corresponding time average $\langle u \rangle$ of each individual prey species in the two-dimensional space. The figure reveals that except at the center where the average amplitude is low (around $2.3$) compared to the remaining space that is occupied with average amplitude of species ranging between $2.5$ to $2.6$. The amplitude difference $\Delta \langle u \rangle$ plotted in Fig. \ref{fig9}(g) along the horizontal cross-section through $y=124$ takes approximately zero value for all the spatial positions except for very few nodes at the center. The frequency of rotation of the spiral wave decreases as one or both the diffusion strengths are increased and an exemplary snapshot of the spatial pattern is presented in Fig. \ref{fig9}(b) for diffusion strengths $D_u = 0.01$ and $D_v = 0.1$. The corresponding finite-time average $\langle u \rangle$ represented in Fig. \ref{fig9}(e) again confirms that the species densities oscillate with nearly identical amplitude throughout the two-dimensional space except at the center. However, the size of the core with drifted amplitude starts to increase as the diffusion strengths increase which is apparent from Fig. \ref{fig9}(h). Finally, Fig. \ref{fig9}(c) in the right panel corresponds to uniform spatial densities of the prey species throughout the two-dimensional space for $D_u = 0.1$ and $D_v = 1.0$. The identical amplitude of oscillating species densities throughout the space is easily recognizable from the Figs. \ref{fig9}(f) and \ref{fig9}(i).

\begin{figure}[ht]
	\centerline{
		\includegraphics[scale=0.45]{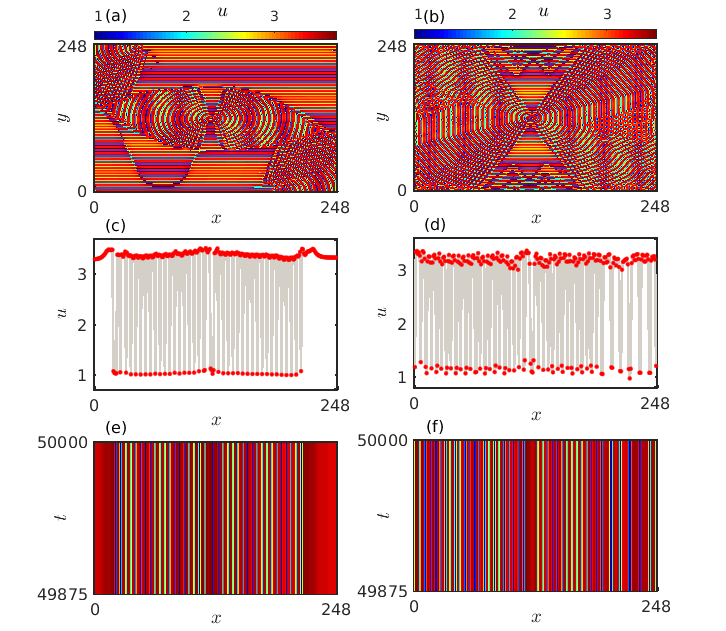}}
	\caption{(a,b) Snapshots of the spatial distribution of the prey species $u$ at a particular time instant $t = 5 \times 10^4$ in two-dimensional diffusive space exhibiting inhomogeneous steady state dynamics, (c,d) corresponding snapshots along the horizontal cross-section through the line $y=125$, (e,f) spatiotemporal behavior of the states presented in (c,d). Diffusion strengths are fixed at $D_u = 0.0018, D_v=1.0$ (blue square in Fig. \ref{fig7}(a)) for the left panel and $D_u = 0.0056, D_v = 1.0$ (cyan square in Fig. \ref{fig7}(a)) for the right panel.}
	\label{fig10}
\end{figure}

Next, we also demonstrate the spatial arrangement as well as  the spatiotemporal organization of the prey densities when they are in the steady state regime. The emergence of Turing instability for suitable combinations of prey and predator diffusion coefficients drives the system to exhibit inhomogeneous steady state dynamics. The transition from amplitude mediated spiral chimeras to inhomogeneous steady states results in two different types of oscillation death states. One is the chimera death state, where a coexistence of coherent and incoherent oscillation death states is observed just like the oscillatory chimera states, and the other is the incoherent oscillation death state, where no coherent behavior is detected among the constituents. Two distinct snapshots of the spatial organization among the prey species in oscillation death state are presented in Figs. \ref{fig10}(a) and \ref{fig10}(b) for diffusion strengths $D_u = 0.0018, D_v =1.0$ and $D_u = 0.0056, D_v = 1.0$, respectively. In the middle panel the corresponding snapshots along the horizontal cross-sections through the line $y=125$ are presented. Coexistence of coherent and incoherent oscillation death states is prominent from Fig. \ref{fig10}(c) which characterizes the state as chimera death state. On the other hand, Fig. \ref{fig10}(d) clearly demonstrates the presence of incoherent oscillation death states which keeps on fluctuating randomly between the upper and the lower branches of the inhomogeneous steady states. The stationary behavior of such states is further confirmed from the spatiotemporal dynamics presented in Figs. \ref{fig10}(e) and \ref{fig10}(f), respectively.

\subsection{Characterization of distinct dynamical states}

In order to characterize the occurrence of all such diverse patterns, we introduce a measure to estimate the strength of incoherent populations in a particular pattern arising in the two-dimensional space. On the basis of the computed finite-time average of the species densities for each pattern, we calculate the strength of incoherent subpopulation for each of those dynamical patterns using the following relation
\begin{equation}
S = 1-\frac{1}{N^2}\sum\limits_x\sum\limits_y H(\bar{u}(x,y)),~~~ H(\bar{u}) = \Theta(\delta-\vert \bar{u} \vert),
\end{equation}
generalizing the notion of strength of incoherence introduced by Gopal et al. \cite{chimera_nonlocal3} to characterize chimera states in one dimensional arrays. Here $\bar{u}(x,y) = \langle u(x,y) \rangle - \langle u_c \rangle$ measures the deviation of each of the  species densities situated at position $(x,y)$ from that of the coherent subpopulations. $\langle u_c \rangle$ represents the average amplitude of the coherent subpopulations throughout the two-dimensional space which is kept fixed at $\langle u_c \rangle =  2.55$. $\bar{u} = 0$ corresponds to the coherent subpopulation whereas the nonzero values of $\bar{u}$ signify the species belonging to the incoherent group. $\delta$ is a predefined threshold value and $\Theta(\cdot)$ denotes the Heaviside step function. 
\begin{figure}[ht]
	\centerline{
		\includegraphics[scale=0.48]{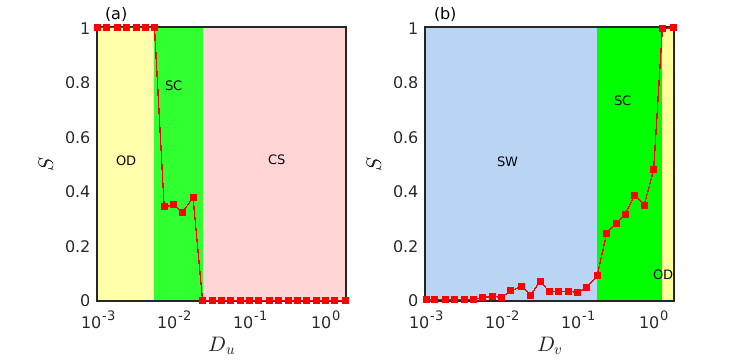}}
	\caption{Strength of incoherent subpopulations $S$ in the RD system \eqref{eq1} (a) for varying prey diffusion rate $D_u$ and fixed predator diffusion rate $D_v = 0.56$, (b) for varying predator diffusion rate $D_v$ and fixed prey diffusion rate $D_u = 0.01$. $\delta=0.035$ is considered here. Presence of four different dynamical states, namely, oscillation death (OD), spiral wave (SW), spiral chimera (SC) and complete synchronization (CS) states are distinguished with yellow, sky blue, green and pink shaded regions, respectively.}
	\label{fig11}
\end{figure}
The measure for the estimation of strength of incoherent subpopulations $S$ lies in the range $[0,1]$. Zero value of $S$ is attained when the system exhibits completely synchronized temporally oscillating uniform spatial densities of the species, while $S=1$ stands for the oscillation death states or the inhomogeneous steady states. Various spiral dynamics are represented by the non-zero and non-unit values of $S$, i.e., $0<S<1$. However, for ordinary spirals due to the presence of very little incoherence around the core, $S$ takes values close to zero. Existence of amplitude mediated spiral chimera states is determined by sufficiently large values of $S$ lying between $0$ and $1$. In Fig. \ref{fig11} the transitions among different dynamical states are illustrated by plotting the measure $S$ with respect to the diffusion coefficients $D_u, D_v$. Three distinct collective dynamics, namely oscillation death (OD), spiral chimera (SC) and completely synchronized oscillation (CS), are observed when the strength of incoherent subpopulations are plotted in Fig. \ref{fig11}(a) with respect to varying prey diffusion rate $D_u$ for fixed value of the predator diffusion rate $D_v = 0.56$. Additionally, in Fig. \ref{fig11}(b) we plot the measure $S$ for varying predator diffusion strength $D_v$ at fixed value of prey diffusion strength $D_u = 0.01$. Here also three distinct dynamical states are detected, namely spiral wave (SW), spiral chimera (SC) and oscillation death (OD) states. From the figure it is evident that the amplitude mediated spiral chimera states appear during the transition from oscillatory motion to steady state dynamics.

\subsection{Effect of spatial geometry}
\begin{figure}[ht]
	\centerline{
		\includegraphics[scale=0.45]{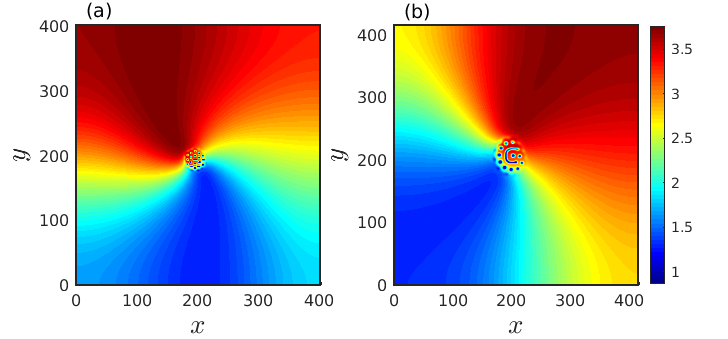}}
	\caption{Effect of grid size $N^2$ and spatial step size $h$ on the spiral chimera dynamics. Snapshots of the spatial patterns are taken at (a) $t=5\times 10^4$ for $N=401, h=0.25$, and $dt= 0.025$, (b) $t = 5.25 \times 10^4$ for $N=415, h=0.15$ and $dt=0.015$. Diffusion coefficients are fixed at $D_u =0.01, D_v=1.0$.}
	\label{fig12}
\end{figure}

Throughout the article, all the numerical simulations are carried out using the fixed grid size $N^2 = 249^2$, spatial step $h=0.25$ and integration time step $dt=0.025$. Now, we check the effect of varying $N$ and $h$ on the observed patterns of the considered system. For this, we fix the diffusion coefficients at $D_u = 0.01$ and $D_v=1.0$ for which spiral chimera pattern was observed in Fig. \ref{fig8} with $N=249$. Now, if the value of $N$ is reduced gradually, the dynamics changes from spiral chimera to completely synchronized oscillation states (figure not shown), whereas increasing $N$ retains the dynamics unchanged as obtained in Fig. \ref{fig8}(a). It should be noted that varying only $N$ alters the spatial distance $N \times h$. Thus, lowering $N$ reduces the spatial distance which may not be sufficient for the emergence of the spiral pattern which is why the dynamics gets changed for smaller values of $N$. Figure \ref{fig12}(a) displays the snapshot of the spatial evolution of the pattern obtained for $N=401$ at a particular time instant $t= 5 \times 10^4$. Obviously, the pattern is same as depicted in Fig. \ref{fig8}(a) since the spatial length $N\times h$ is sufficiently large here. On the other hand, we have also verified that similar spatial patterns can be observed by varying the spatial step size $h$, but the time step $dt$, time iteration and $N$ should also be changed accordingly. For example, we choose $h=0.15, dt = 0.015, N=415$ and time iteration $= 3.5 \times 10^6$  such that the spatial distance $N\times h$ and the time $t$ (iteration$ \times dt$) at which the snapshot is recorded remains almost unchanged. The snapshot of the spatial pattern is depicted in Fig. \ref{fig12}(b). Therefore, other choices of $N, h, dt$ may also give rise to the similar dynamics as long as the spatial distance $N\times h$ and the time iteration $t/dt$ remains sufficient for the emergence of the respective patterns.

\section{Conclusion} \label{sec6}
In summary, through this article, we have established a connection between the different instabilities, namely Turing and Hopf instabilities that are responsible for pattern formation in spatially extended reaction-diffusion systems and the emergence of chimera patterns in such systems. Unfolding of diverse collective dynamics including the exotic chimera states has been reported by considering a two component prey-predator model over a two-dimensional diffusive medium. We have performed a detailed linear stability analysis of the considered RD system to derive the conditions for both the onset of Turing and Hopf bifurcations which enable us to explore the parametric region where the stability of the homogeneous steady state of the system is lost. Emergence of coherent spiral waves has been detected in a broad region of the diffusion driven parameter space where the system exhibits temporal oscillations beyond the Hopf bifurcation. Besides numerical simulations, development of this spiral pattern has been verified analytically by deriving the appropriate linear amplitude equation (in Appendix \ref{appendix}). The most interesting phenomenon that we have observed is the appearance of random fluctuations around the spiral core for parameter values exceeding both the Turing and Hopf bifurcations. Simultaneous interplay of the oscillatory dynamics formed due to Hopf bifurcation and the emergence of inhomogeneous steady states caused by the Turing bifurcation combinedly generates a randomized core inside the coherent spiral arm. Such coexistence of coherent-incoherent dynamics within a spiral structure is represented as spiral chimera pattern and their presence is manifested for suitable combinations of the prey and predator diffusion coefficients in the parametric space. 

Previously, spiral chimera states were observed in a similar model \cite{sw_epjst} in the absence of intrapredator interaction term, where the coherent-incoherent dynamics was observed only on the phases of the oscillators. In contrast to this, the inclusion of the term $\eta v^2$ in our current study develops an amplitude mediated spiral chimera displaying randomness both in phases and amplitudes of the constituents. We confirmed the occurrence of such states by looking into the finite-time averages and the phases of the individual dynamical units. In fact, the additional term also helped in the analytical derivation of the linear amplitude equation corresponding to the spiral pattern presented in Appendix \ref{appendix}. Instead of incoherent oscillatory states observed in \cite{sw_epjst}, here we obtained various oscillation death states owing to the amplitude variations. Concurrence of both Hopf and Turing instabilities promotes the emergence of diverse dynamical states, precisely, spiral wave, spiral chimera, synchronized oscillation and oscillation death states. We found that the amplitude mediated spiral chimera states appear during the transition from temporally oscillating states to inhomogeneous steady states with respect to varying diffusion coefficients. The occurrence of all these states are characterized by defining a measure $S$, to estimate the strength of incoherent subpopulations corresponding to a particular pattern. Moreover, we also verified that the spiral chimera pattern emerges for other appropriate choices of $N, h$ and $dt$. Previous studies on the emergence of spiral chimera states either in RD or non-RD systems have revealed the appearance of incoherent dynamics only in the phases of the constituting elements. Surprisingly, the emergence of an amplitude mediated spiral chimera state revealing incoherent behavior both in terms of the phases and the amplitudes of the composing units have not been reported earlier. Such fascinating dynamics arises for parameter values surpassing both the Turing and Hopf bifurcations, which we believe to be an important observation of our current study.

In ecology, dispersal of species in terms of diffusion not only helps in maintaining species diversity and species persistence, but also supports various self-organization phenomena associated with species invasion, colony formation or extinction in ecosystems. Ref. \cite{drake} suggests that the synchronization in coupled ecological dispersal networks might get disrupted by species invasion which can be related with the occurrence of chimera like behavior. 

\par {\bf Acknowledgments:}
The work of P.M. is supported by DST-FIST (Department of Physics), DST-PURSE and MHRD RUSA 2.0 (Physical Sciences) Programmes. M.L. is supported by Science and Engineering Research Board Distinguished Fellowship.

\appendix

\section{Amplitude equation for spiral pattern} \label{appendix}

For further understanding on the development of spiral pattern beyond the Hopf bifurcation boundary, here, we proceed through the derivation of the amplitude equation for spiral pattern following a multiscale analysis procedure \cite{multiscale1, multiscale2}. 
For this, we take the linearized system Eq. $\eqref{eq7}$ to find the amplitude equation. For our system, we consider $a$ to be the bifurcation parameter and $a_0 $ is the value of $a$ at which Hopf bifurcation occurs, as obtained in Sec.\ref{temporal}. We can tune the control parameter as $a = a_0 + \epsilon a_1$ close to the bifurcation boundary for the spiral wave solution, where $\epsilon$ is the smallness parameter such that $0 < \epsilon \ll 1$. Now, we scale the two-dimensional space coordinates as $r = r_0 + \epsilon r_1$ and $\theta = \theta_0 + \sqrt{\epsilon}\theta_1$ and the time variable as $t = t_0 + \epsilon \tau$, which give
\begin{equation}
\begin{array}{lcl}
\frac{\partial}{\partial r} = \frac{\partial}{\partial r_0} + \epsilon \frac{\partial}{\partial r_1},~~
\frac{\partial}{\partial \theta} = \frac{\partial}{\partial \theta_0} + \sqrt{\epsilon} \frac{\partial}{\partial \theta_1},~~
\frac{\partial}{\partial t} = \frac{\partial}{\partial t_0} + \epsilon \frac{\partial}{\partial \tau}.
\end{array}
\end{equation}
The perturbation terms $\tilde{u}$ and $\tilde{v}$ can be expressed as 
\begin{equation}
\begin{array}{lcl}\label{eq16}
\tilde{u} = \epsilon u_1 + \epsilon^2 u_2 + \hdots,\\
\tilde{v} = \epsilon v_1 + \epsilon^2 v_2 + \hdots.
\end{array}
\end{equation}
Substituting the multiple scale expansion given by Eq. \eqref{eq16} along with the rescaled coordinates into \eqref{eq7} we get,
\begin{equation}\label{eq17}
\begin{array}{lcl}
\epsilon \Big[\frac{\partial u_1}{\partial t_0} - a_0 (f_u u_1+ f_v v_1)- D_u \nabla_0^2u_1 \Big] \\+ \epsilon^2 \Big[\frac{\partial u_1}{\partial \tau}+\frac{\partial u_2}{\partial t_0} -a_0(f_u u_2 + f_v v_2) -a_1(f_u u_1+ f_v v_1) \\~~~~~~ - D_u(\nabla_0^2u_2 + \nabla_1^2 u_1) \Big] = 0,
\end{array}
\end{equation}
and
\begin{equation}\label{eq18}
\begin{array}{lcl}
\epsilon \Big[\frac{\partial v_1}{\partial t_0} - (g_u u_1+ g_v v_1)-D_v\nabla_0^2v_1 \Big] \\+ \epsilon^2 \Big[\frac{\partial v_1}{\partial \tau}+\frac{\partial v_2}{\partial t_0} - (g_u u_2 + g_v v_2) -D_v(\nabla_0^2v_2 + \nabla_1^2 v_1)\Big] = 0.
\end{array}
\end{equation}
Here the Laplacian $\nabla^2$ is expanded as 
\begin{equation}\label{eq19}
\nabla^2 = \nabla_0^2 + \epsilon \nabla_1^2,
\end{equation}
considering terms only up to $O(\epsilon)$, where 
\begin{equation}
\begin{array}{lcl}\label{eq20}
\nabla_0^2 = \Big[\frac{1}{r_0}\frac{\partial}{\partial r_0} (r_0\frac{\partial}{\partial r_0}) + \frac{1}{r_0^2} \frac{\partial^2}{\partial \theta_0^2}\Big], \\\\
\nabla_1^2 = \Big[\frac{1}{r_0}\frac{\partial}{\partial r_0}(r_0\frac{\partial}{\partial r_1}) + \frac{1}{r_0^2} \frac{\partial^2}{\partial \theta_1^2}\Big].
\end{array}
\end{equation}
Collecting the $O(\epsilon)$ terms from Eqs. \eqref{eq17} and \eqref{eq18} we get,
\begin{equation}\label{eq21}
\bf L \begin{pmatrix}
u_1 \\ v_1
\end{pmatrix} = \begin{pmatrix}
0 \\ 0
\end{pmatrix},
\end{equation}
where
\begin{equation}\label{eq22}
\bf L = \begin{pmatrix}
\frac{\partial}{\partial t_0}-a_0 f_u - D_u\nabla_0^2 & -a_0 f_v \\ -g_u & \frac{\partial}{\partial t_0} - g_v - D_v \nabla_0^2
\end{pmatrix}.
\end{equation}
Assuming the trial solution of the following form,
\begin{equation}\label{eq23}
\begin{pmatrix}
u_1 \\ v_1
\end{pmatrix} = A(r_1, \theta_1, \tau)
\begin{pmatrix}
\tilde{u}_1 \\ \tilde{v}_1
\end{pmatrix} e^{\lambda t_0},
\end{equation}
we obtain the eigenvalue $\lambda$ in the absence of diffusion as given by Eq. \eqref{eq5}. Eigenvalues become purely imaginary at the Hopf boundary $a_0$ and the Hopf frequency is given by $\omega_H = \sqrt{a_0(f_u g_v- f_v g_u)}$. 

Substituting Eq. \eqref{eq23} into \eqref{eq21} we obtain $\frac{\tilde{u}_1}{\tilde{v}_1} = \frac{\lambda-g_v}{g_u}$, which yields the explicit solutions of $u_1$ and $v_1$ as
\begin{equation}\label{eq24}
\begin{pmatrix}
u_1 \\ v_1 
\end{pmatrix} 
= A(r_1,\theta_1,\tau)
\begin{pmatrix}
\lambda-g_v \\ g_u 
\end{pmatrix} e^{\lambda t_0} + c.c.,
\end{equation} 
where $c.c.$ stands for the complex conjugate of the preceding term.

Now, we solve the equation for the next order, $O(\epsilon^2)$. Collecting the coefficients from Eqs. \eqref{eq17} and \eqref{eq18}, we obtain the coupled equations
\begin{equation}\label{eq25}
\bf L \begin{pmatrix}
u_2 \\ v_2
\end{pmatrix}
= \begin{pmatrix}
-\frac{\partial}{\partial \tau}+a_1 f_u + D_u\nabla_1^2 & a_1 f_v \\ 0 & -\frac{\partial}{\partial \tau}+ D_v\nabla_1^2
\end{pmatrix}
\begin{pmatrix}
u_1 \\ v_1
\end{pmatrix}.
\end{equation}
Since, the operator $\bf L$ has zero eigenvalue corresponding to the eigenstate $\begin{pmatrix}
u_1 \\ v_1
\end{pmatrix}$, we must apply Fredholm solvability condition to Eq. \eqref{eq25} for the existence of a solution. Accordingly, the R.H.S. of Eq. \eqref{eq25} and the left eigenvector of $\bf L$ associated with the zero eigenvalue must be orthogonal, which gives 
\begin{equation}\label{eq26}
\begin{pmatrix}
u_1^\dagger & v_1^\dagger
\end{pmatrix}
\begin{pmatrix}
-\frac{\partial}{\partial \tau}+a_1 f_u + D_u\nabla_1^2 & a_1 f_v \\ 0 & -\frac{\partial}{\partial \tau}+ D_v\nabla_1^2
\end{pmatrix}
\begin{pmatrix}
u_1 \\ v_1
\end{pmatrix} = 0,
\end{equation}
where `$\dagger$' denotes the complex conjugate.
Substituting Eq. \eqref{eq24} into \eqref{eq26}, we obtain,
\begin{equation}\label{eq27}
\frac{\partial A}{\partial \tau} =  C_1 \nabla_1^2 A + a_1 (C_2 + i C_3 )A, 
\end{equation}
where $C_1 = \frac{D_u(\omega_H^2+g_v^2) + D_v g_u^2}{g_u^2+g_v^2+\omega_H^2}$, $C_2 = \frac{f_u(\omega_H^2+g_v^2) - g_u f_v g_v}{g_u^2+g_v^2+\omega_H^2}$ and $C_3 = \frac{f_v g_u\omega_H}{g_u^2+g_v^2+\omega_H^2}$. Transforming back to the original scales finally we obtain the desired amplitude equation,
\begin{equation}\label{eq28}
\frac{\partial A}{\partial t} = C_1\Big(\frac{1}{r}\frac{\partial A}{\partial r} + \frac{1}{r^2} \frac{\partial^2 A}{\partial \theta^2}\Big) + \epsilon a_1C_2A + i\epsilon a_1C_3A.
\end{equation}
The derived equation now can be solved to obtain spirals for which a general solution is assumed of the following form,
\begin{equation}\label{eq29}
A(r,\theta,t) = \tilde{A}(r)e^{i(\omega t + \psi(r) + m\theta)}.
\end{equation}
The distance of any point from the core is given by $r$ and the polar angle around the core is $\theta$. $\tilde{A}(r)$ and $\psi(r)$, respectively represent the amplitude and phase of the spiral wave, $\omega$ is the wave frequency and $m$ is the spiral-arm number.

Substituting the general solution \eqref{eq29} into \eqref{eq28} and equating the coefficients of the real and imaginary parts, we obtain
\begin{equation}
\begin{array}{lcl}\label{eq30}
\frac{C_1m^2}{r^2} = \frac{C_1}{r\tilde{A}}\frac{d\tilde{A}}{dr} + \epsilon a_1 C_2, \\\\
\omega = \frac{C_1}{r}\frac{d\psi}{dr} + \epsilon a_1C_3.
\end{array}
\end{equation}
Now, solving Eq. \eqref{eq30} for $\tilde{A}$ and $\psi$ we derive the approximated expression for spiral pattern as,
\begin{equation}\label{eq31}
\begin{array}{lcl}
\tilde{A}(r) = I_1r^{m^2}e^{-\epsilon a_1C_2r^2/2C_1},\\
\psi(r) = \frac{\omega - \epsilon a_1C_3}{C_1} \frac{r^2}{2} + I_2,
\end{array}
\end{equation}
where $I_1$ and $I_2$ are integration constants.

\begin{figure}[ht]
	\centerline{
		\includegraphics[scale=0.45]{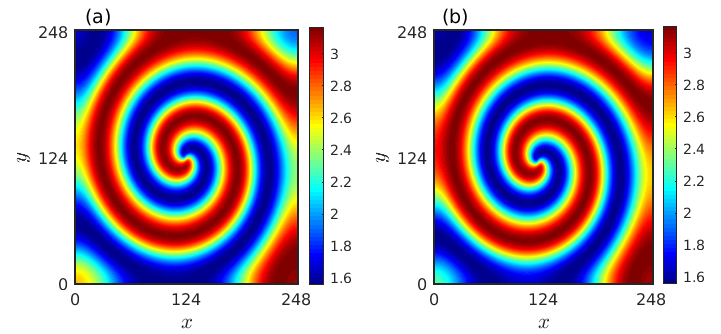}}
	\caption{Verification of the numerically simulated spiral pattern with the analytically derived solution. Parameter values are fixed at $a=2.0, D_u = 0.01, D_v = 0.001$. (a) Spiral pattern obtained directly from numerical simulation, (b) spiral obtained by taking the approximate expression \eqref{eq32} as the initial condition for $I_1 = I_2 = 1.0$.}
	\label{fig13}
\end{figure}

For validation of the obtained solutions we consider as an illustrative example the choice $m =1, \epsilon = 0.01, a_1 = 27.21$ and calculate $C_1 = 3.06\times10^{-3}, C_2 = 4.493\times10^{-7}, C_3 = -0.0635, \omega_H = 0.4788, a_0 = 1.7279$ for the chosen parameter values and we find 
\begin{equation}\label{eq32}
\begin{array}{lcl}
\tilde{A}(r) = I_1re^{-\epsilon a_1(7.342\times 10^{-5})r^2},\\\\
\psi(r) = \frac{\omega + 0.0635 \epsilon a_1}{6.12\times10^{-3}}r^2 + I_2.
\end{array}
\end{equation}
As the spiral pattern in the reaction-diffusion system emerges due to the occurrence of Hopf bifurcation, the frequency $\omega$ for the spiral pattern is controlled by the Hopf frequency $\omega_H$. So, we choose $\omega$ close to $\omega_H$ and taking specific values of $t, I_1, I_2$, we get $O(\epsilon)$ approximated solution as $u = u_s + \epsilon Re(u_1), v = v_s + \epsilon Re(v_1)$. We consider this approximate expression as the initial condition and perform numerical simulation of the model \eqref{eq1}. The resultant spiral pattern as depicted in Fig. \ref{fig13}(b) closely matches with the spiral pattern presented in Fig. \ref{fig13}(a) that is obtained directly from numerical simulation.

\end{document}